\documentclass{iaus}
\usepackage{graphicx}

\title[Star clusters] 
{The star clusters of the Magellanic System}

\author[Santiago]   
{Bas\'\i lio Santiago$^1$
}

\affiliation{$^1$Instituto de F\'\i sica, Universidade Federal do Rio Grande do Sul \\ Caixa Postal 15051, Porto Alegre, RS, Brazil \\ email: {\tt santiago@if.ufrgs.br} \\[\affilskip]
}

\pubyear{2008}
\volume{256}  
\jname{The Magellanic System: Stars, Gas, and Galaxies}
\editors{Jacco Th. van Loon \& Joana M. Oliveira, eds.}
\begin{document}

\maketitle

\begin{abstract}
More than 50 years have elapsed since the first studies of star clusters 
in the Magellanic Clouds. The wealth of data accumulated since then has not 
only revealed a large cluster system, but also a diversified one, filling loci 
in the age, mass and chemical abundance parameter space which are 
complementary to Galactic clusters. Catalogs and photometric samples 
currently available cover most of the cluster mass range. The expectations of 
relatively long cluster disruption timescales in the Clouds have been 
confirmed, allowing reliable assessments of the cluster initial mass function 
and of the cluster formation rate in the Clouds. Due to their proximity to the 
Galaxy, Magellanic clusters are also well resolved into stars. Analysis of 
colour-magnitude diagrams (CMDs) of clusters with different ages, masses and 
metallicities are useful tools to test dynamical effects such as mass loss 
due to stellar evolution, two-body relaxation, stellar evaporation, cluster 
interactions and tidal effects. The existence of massive and young Magellanic 
clusters 
has provided insight into the physics of cluster formation. The magnitudes and 
colours of different stellar types are confronted with stellar evolutionary 
tracks, 
thus constraining processes such as convective overshooting, stellar 
mass-loss, rotation and pre main-sequence evolution. Finally, the Magellanic 
cluster system may contribute with nearby and well studied counterparts of 
recently proposed types of extragalactic clusters, such as Faint Fuzzies and 
Diffuse Star Clusters.
\end{abstract}

\firstsection 

\section{The census of star clusters in the Magellanic Clouds}

\begin{figure}[b]
\begin{center}
 \includegraphics[width=2.3in]{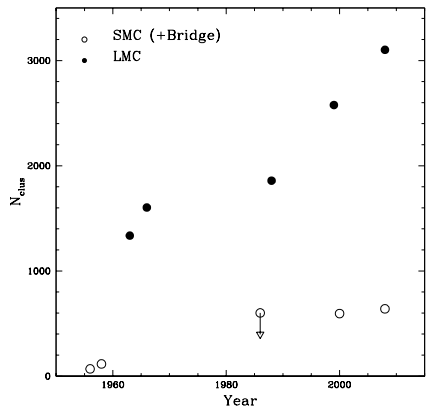} 
 \caption{Census of known star clusters as a function of time. LMC: solid 
circles; SMC and Bridge: open circles.}
   \label{cluscensus}
\end{center}
\end{figure}

The first papers on star clusters in the Magellanic
Clouds (MCs) date from over 50 years ago. They were concerned
with detecting these objects and measuring their basic
properties such as positions, apparent sizes and position angles.
Figure \ref{cluscensus} shows the evolution in the census of
star clusters in the Clouds from these early papers to the
present. The numbers shown were taken from 
references in the literature spanning the entire period 
(Kron 1956, Lindsay 1958, Shapley \& Lindsay 1963, Lynga \& Westerlund
1963, Hodge \& Sexton 1966, Hodge 1986, Hodge 1988, Bica et al 1999,
Bica \& Dutra 2000, Bica et al 2008).
A steady increase in the number of known clusters can
be seen, especially in the case of the Large Magellanic
Cloud (LMC). For the Small Magellanic
Cloud (SMC), the rise has been slower.
This may, at least in part, reflect the fact that some of the earlier
catalogs did not separate star clusters from stellar associations.

\begin{figure}[b]
\begin{center}
 \includegraphics[width=2.5in]{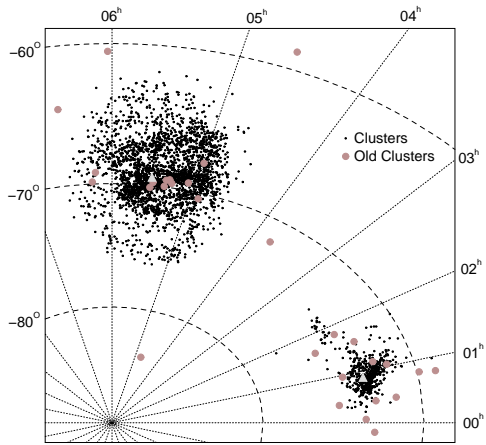} 
 \caption{On-sky distribution of the 3740 star clusters in the Bica et al (2008) catalogue. A grid of equatorial coordinates is also shown. The larger dots are the clusters with estimated ages $\tau > 4 Gyrs$. }
   \label{onsky}
\end{center}
\end{figure}

\begin{figure}[t]
\begin{center}
 \includegraphics[width=2.5in]{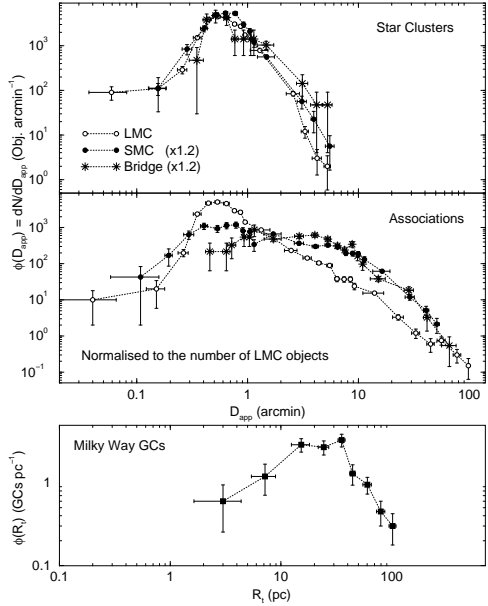} 
 \caption{Top panel: apparent diameter function of star clusters of 
Bica et al (2008). The LMC, SMC and Bridge clusters are shown in separate,
as indicated. Middle panel: The diameter function of the 3326 stellar 
associations, again separately showing the LMC, SMC and Bridge systems. 
Lower panel: The distribution of tidal radii of Galactic globular clusters.}
   \label{diamfunc}
\end{center}
\end{figure}

The latest catalog has just recently been published by 
Bica et al (2008). It includes star clusters, associations
and other extended objects. It contains a total of 9305 entries,
3740 of which have been classified as star clusters. The on-sky
distribution of star clusters in Bica et al (2008) is shown
in Figure \ref{onsky}. We see that the main structural components
of the Magellanic System are traced by these objects,
including the LMC bar and outer ring, the SMC wing and the bridge.

One important issue is whether the current census of stars
clusters is complete in the MCs. A bias against faint 
and/or compact star clusters should exist to some extent,
as these objects are harder to detect or to separate from 
single bright stars. The apparent diameter function
of star clusters from Bica et al (2008) is shown in Figure \ref{diamfunc}, 
separately
for the LMC, SMC and Bridge clusters. The clusters in the SMC and Bridge
have been scaled to the LMC distance. The diameter
function shows a power law behaviour at large diameters,
reaching a peak at $D_{app} \simeq 0.6'$ and then dropping sharply
at the small diameter end. The distribution of
tidal radii of Galactic globular clusters is also shown
in the figure and has a similar shape. As these latter should
make up an essentially complete sample, the authors conclude that this
down turn is a real feature, rather than the result of
a selection bias. 

\section{Basic properties of the MC star cluster system}

Since the LMC and SMC are at distances of $\simeq$ 50 kpc and 60 kpc,
respectively, their system of star clusters can be studied 
in much more detail than those of more distant hosts. 
For instance, the photometric sample of LMC clusters by Hunter et al
(2003) is complete down to $M_V \simeq -3.5$. This
is 4 magnitudes fainter than the absolute magnitude
at the peak of the globular cluster luminosity function (GCLF)
observed in other galaxies, especially in luminous early-type 
ones. Therefore, the LMC and SMC cluster systems are probed towards
much lower masses than elsewhere. 

Basic properties of hundreds of individual star clusters in both
Clouds, such as ages, metallicities 
and masses, have been obtained from integrated photometry, 
integrated spectroscopy or colour-magnitude diagrams (CMDs) resulting
from high-resolution imaging.

\begin{figure}[t]
\begin{center}
 \includegraphics[width=2.5in]{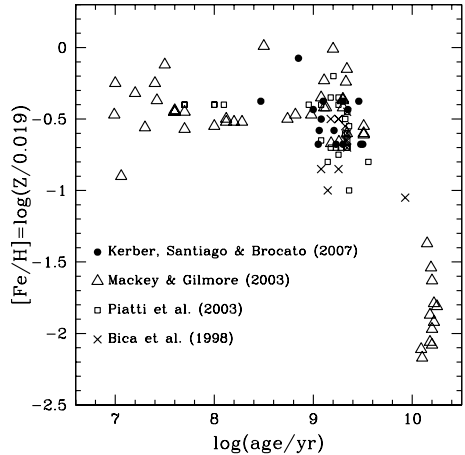} 
 \caption{The age-metallicity relation for LMC clusters, taken from Kerber et 
al (2007). The different symbols show clusters from different samples,
as indicated.}
   \label{LMC-AMR}
\end{center}
\end{figure}

The age-metallicity relation (AMR) of rich LMC clusters 
compiled by Kerber et al (2007) is shown in Figure \ref{LMC-AMR}. It 
allows us to discuss the main properties of the LMC cluster system.
Rich clusters span a wide range in ages, 
$7 \leq log \tau(yrs) \leq 10$. Most clusters younger than 
$\simeq 3 Gyrs$ have metallicities between half and one third of 
the solar value. Only about 15 rich clusters are older than $10 Gyrs$ and have 
$[Fe/H] \simeq -1.5$ or less, thus having properties similar
to those of globular clusters (GCs) found in the Galactic halo. 
A noticeable feature is the so-called {\it age gap}, as only
one cluster, ESO121SC03, is found in the range $3 Gyrs \leq \tau \leq
10 Gyrs$. This cluster has recently been studied by Xin et al (2008),
using Hubble Space Telescope Advanced Camera for Surveys (HST/ACS), 
and its age has been confirmed to fall in this interval.

The SMC cluster system is also rich and diversified in terms of
ages and abundances. A recent AMR for the SMC can be found in 
Piatti et al (2007) (see also contribution by E. Grebel to these
proceedings). Despite the often large error bars in determined
cluster ages, the SMC seems 
to have several examples of intermediate-age clusters, with no clear sign
of an age gap. Another difference relative to the LMC is
the absence of a significant population of rich, old and metal poor
clusters similar to the Galactic GCs. Only NGC 121 has GC properties
(Glatt et al 2008).

\section{Age and mass distributions: constraining the cluster formation rate and initial mass function}

Large photometric surveys of both clouds are currently 
available, yielding magnitudes and colour for millions of stars and
hundreds of star clusters (Massey 2002, Zaritsky et al 1997,
Hunter et al 2003, Rafelski \& Zaritsky 2005).

The integrated magnitudes and colours of star clusters can be
compared to those of single stellar populations, as predicted
by different models (Leitherer et al 1999, Kurth et al 1999).
As a result, age and mass estimates can be inferred for each
cluster. The distribution of clusters with these quantities
bears a lot of information on the cluster initial mass function
(ICMF), the cluster formation history (CFR) and the
timescale for cluster disruption.

As an example, Boutloukos \& Lamers (2003) have modelled
the age distribution of star clusters.
The model assumes that the ICMF is a power-law with cluster mass,
$\propto m^{-2}$, and that the CFR has been uniform.
Assuming also that the cluster disruption timescale, $t_d$, is a power-law as
a function of cluster mass, the model predicts that the number
of star clusters per unit age should be described by a double power-law
as a function of age , with
a shallow slope at small ages and a steeper one for the older
clusters. The slower drop at small ages results from fading:
clusters become less luminous as they age, so that their number decreases in
any magnitude limited sample. At larger ages, dynamical evolution
leads to the disruption of less massive clusters, leading to a faster
drop in the observed age distribution. This behaviour was in fact observed
by Boutloukos \& Lamers (2003) in several galaxies with
sizable cluster samples with age estimates, including the SMC.
From the age distribution of the SMC clusters, they infer
a disruption timescale of 

$$t_{d} \simeq 8 \Bigl ( { {M} \over {10^4 M_{\odot}} } \Bigr )^{0.62}~ Gyrs$$  

Hunter et al (2003) have obtained integrated magnitudes and colours
for 939 SMC and LMC clusters using the images from the Massey (2002)
survey. They have modelled the age and mass distributions and found
evidence for cluster fading as well as an increase in maximum cluster mass for
larger ages resulting from statistical sampling, from which they
were able to constrain the ICMF slope.

de Grijs \& Anders (2006) re-analyzed the Hunter et al (2003) data
using a tool to compare clusters colours and ages to SSPs and applied
the Boutloukos \& Lamers (2003) model in order to infer 
a $t_d = 8$ Gyrs for $10^4~M_{\odot}$ LMC clusters, in
agreement with the results for the SMC.

In brief, these studies quantitatively confirm that star clusters tend 
to live much longer in the Clouds than in the Galaxy as a result of 
slower disruption processes. We point out that, qualitatively, the 
differences in age distribution between the Galaxy and the Clouds is long 
known (Hodge 1988). These studies based on integrated photometry 
also constrained the ICMF slope and confirmed a roughly constant CFR in 
the MCs, apart from the age gap in the LMC. 

\section{Magellanic star clusters in high resolution: structure and dynamics}

The HST has allowed individual star clusters to be resolved
into stars and studied in detail (Brocato et al 2001, Santiago et al 2001). 
Detailed and self-consistent modelling of the colour-magnitude diagrams 
provides a strong tool for determining the physical properties of rich star
clusters. Self-consistency should be understood in this context as the
capacity to extract the relevant parameters, such as age, metallicity,
distance, extinction and mass functions at different positions, 
all from the same data-set, without pre-fixing any of them.
This may often be the most reliable approach considering the spatial
depth of the two Clouds, the granularity of the dust distribution and
the range in metallicities in these galaxies.
Kerber et al (2002) and Kerber \& Santiago (2005) presented a 
statistical method, 
based on a code from D. Valls-Gabaud, to model 
observed CMDs by comparing them to synthetic ones. The authors 
applied this tool to several LMC star clusters observed with HST's Wide
Field and Planetary Camera 2 (HST/WFPC2).
Similar techniques are also successfully applied to CMDs of field stars 
in the Clouds and other dwarf satellites of the Galaxy, 
in order to reconstruct the star formation history
(SFH) in them (Hernandez et al 2000, Javiel et al 2005, No\"el et 2007).

The MCs are an excellent laboratory to observe and constrain
dynamical effects on stellar systems, in special star clusters.
Dynamical and structural evolution can be observed and modelled 
from analysis of high resolution images in LMC and SMC clusters.

First, clusters respond to the evolution of their stars. 
Strong winds from massive stars and supernovae (SN) explosions lead to
mass loss and expansion in young clusters, as observed by Bastian \& 
Goodwin (2006). Any gas not converted into stars before
the first SN bursts will be swept out from the cluster. If the
efficiency of star formation is low, mass loss can be severe,
the remaining cluster becoming unbound and dispersing into
the field at its early stages of evolution. This process is 
usually referred to as {\it infant mortality}.
In fact, the infant mortality rate of Magellanic clusters, especially
in the SMC, has been subject to a lot of recent controversy. Based on
the decline in the number of clusters as a function of age, Chandar
et al (2006) have suggested that up to
$90\%$ of the young star clusters formed in the SMC do not survive the
early stages. Kruijssen \& Lamers (2007) also needed a similarly
large infant mortality rate to accommodate the observed 
age distributions of SMC clusters and field stars under the same general SFH.
On the other hand, Gieles et al (2007) and de Grijs \& Goodwin 
(2008) interpret the declining number of SMC clusters at young ages
as being dominated by the effect of cluster fading in brightness in a 
magnitude-limited sample. More on infant mortality is discussed by R. de Grijs
in these proceedings.

\begin{figure}[b]
\begin{center}
 \includegraphics[width=2.8in]{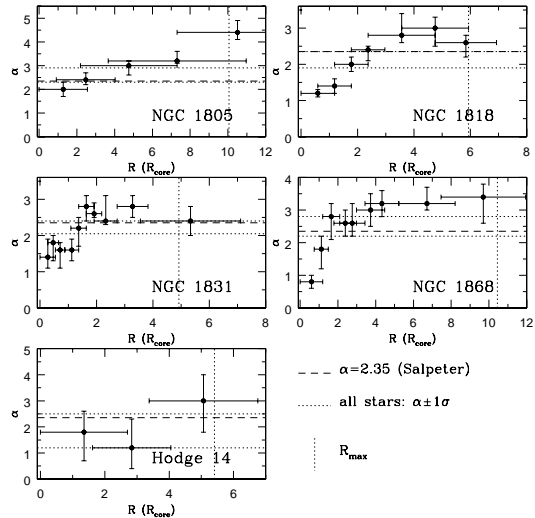} 
 \caption{Variation in PDMF slope as a function of distance from the cluster 
centre for a sample of rich LMC clusters imaged with HST/WFPC2 and studied
by Kerber \& Santiago (2006) and de Grijs et al (2002). Each panel 
shows the slope $\alpha~vs~R$ relation for a different cluster, as 
indicated. The distances are in units of the core radius. The dashed line
corresponds to $\alpha=2.35$ and the horizontal dotted lines indicate 
the $1\sigma$ range around the global PDMF slope.}
   \label{alphaXR}
\end{center}
\end{figure}

Second, two-body interactions among stars occur inside clusters,
as the relaxation timescale is shorter than a Hubble time and
may in fact be of a few Myrs in compact and poor clusters.
As a result of these interactions, more massive stars will 
tend to donate part of their orbital energy to lower mass ones; the
former will tend to sink towards the gravitational potential well,
whereas the latter will achieve less bound orbits, causing
mass segregation inside the cluster. A fraction of the low mass
stars will acquire escape velocity and leave the cluster (stellar
evaporation).

Evidence for mass segregation and stellar evaporation 
is often observed in rich star clusters for which deep and
high-resolution photometry is available, including young ones. 
Figure \ref{alphaXR} shows that the present day mass function (PDMF) slopes
of several LMC clusters modelled by Kerber \& Santiago (2006)
increase steadily as a function of distance from the cluster centre. 
The slopes $\alpha$ result from power-law fits to the
differential number of stars as a function of initial stellar 
mass, $dN/dm \propto m^{-\alpha}$,
in different radial annuli around the cluster centre.
At the largest distance bins, the trend flattens out or is even reversed
in some cases, due to the loss of evaporating low mass stars from the cluster.
Notice also that the global PDMF slopes are consistent with the Salpeter value
($\alpha = 2.35$). 

For young enough clusters, the PDMF should reflect more closely the
initial mass function (IMF). In particular, as the MCs are
metal poorer than the Galaxy, and have a large
system of young and rich clusters, mass function analysis in these 
clusters may help constrain variations in the IMF with
environment. Kumar et al (2008) have
analyzed the mass function of 9 young star clusters in the LMC and found
power-law slopes (in the mass range $2 < m/m_{\odot} < 12$) 
consistent with one another for all but one of them. The slopes are 
also close to the Salpeter value, supporting the idea of a universal IMF.
Schmalzi et al (2008) have also found a nearly Salpeter value for 
the IMF slope in the SMC star forming regions NGC 602, in the
mass range $1 \leq m/m_{\odot} \leq 45$.
On the other hand, at lower masses ($m/m_{\odot} < 1$), the existence
of a universal IMF is still in debate (see contribution by D. Gouliermis in
these proceedings).

\begin{figure}[b]
\begin{center}
 \includegraphics[width=3.5in]{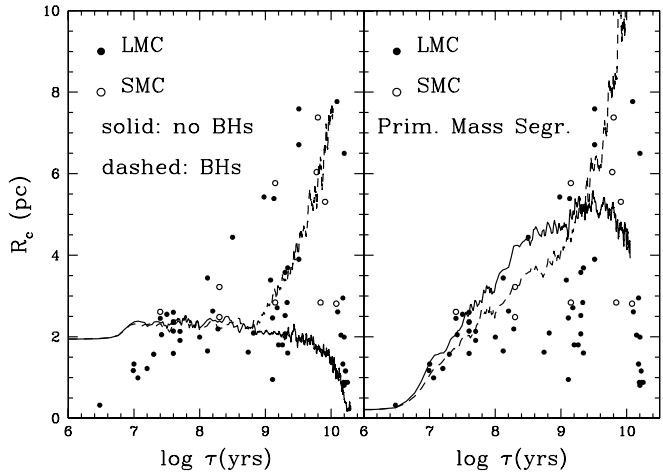} 
 \caption{Left panel: Core radius ($R_c$, in parsecs) as a function of age 
($log \tau(yrs)$) for LMC and SMC star clusters. The data are from Mackey
et al (2003a,b). The lines show the expected evolution based on 
N-body simulations.
Solid line: no heating from stellar degenerates; dashed line: heating included.
Right panel: the data are the same as in the previous panel. The simulation
results now include primordial mass segregation, again with and
without heating by stellar black holes.}
   \label{rcxage}
\end{center}
\end{figure}

A clear evidence of structural evolution of star clusters
is given by the relation between core size ($R_c$) and age. 
The core radius is determined by fitting an equilibrium
model, such as the Elson et al (1987) model, to the cluster surface brightness 
profile. The relation between $R_c$ and age, using the data on rich LMC and SMC
clusters from Mackey et al (2003a,b), is shown in Figure \ref{rcxage}.
The mean core radius clearly increases with age, as does the spread
around the mean. This shows not only that clusters change their
structure with time but also that they do not follow a single path of
core evolution. In fact, a large fraction of the clusters older than
10 Gyrs shown in the Figure have very small cores, possibly having
undergone core collapse. 

Several mechanisms have been proposed to account for core evolution.
Elson et al (1989) explored the possibility that variations in the IMF 
from one cluster to another could lead to different rates of mass loss
and core expansion, therefore explaining the spread in the $R_c$-age relation.
The mean trend towards larger cores has been tentatively explained
as a sampling effect (Hunter et al 2003),
or as the result of dynamical heating of the core by binary stars (Wilkinson
et al 2003) or by stellar black holes (Mackey et al 2008).

Mackey et al (2008) used realistic N-body simulations of star
clusters, in which the number of particles in the simulation is
comparable to the number of real stars. They were able to reproduce
the mean trend in core size with age by including the effect of
core heating by black holes that sink to the cluster centre.
The core evolution in simulations with and without the effect is
shown in the left panel of Figure \ref{rcxage}. The upper boundary
of the observed relation in LMC and SMC clusters can be reproduced
for clusters with primordial mass segregation (right panel), 
especially in the case involving heating by black holes.

\section{MC star clusters as probes to stellar evolutionary theory}

Compared to the Galaxy, the Magellanic Clouds are gas rich and metal poor, 
thus providing a different environment where stars form and evolve. As such, 
they are a complementary and very useful laboratory to test stellar 
evolutionary theories. Observed cluster CMDs are particularly useful
as they can be approximated as single stellar populations and compared to
model isochrones. We here quote some recent examples focussed on very different
stellar evolutionary phases.

Mucciarelli et al (2007a,b) analyzed HST/ACS CMDs of two rich intemediate-age
star clusters in LMC, namely NGC 1978 and NGC 1783.
They compared several model expectations, with convective regions
that exceed the classical one by values in the range $0 \leq \Lambda \leq 
25\%$,
to their high-resolution CMDs. As a result, they favour a mild or large
amount of convective overshooting ($\Lambda = 0.10-0.25$ ) in the
intermediate mass stars at the main sequence turn-off, sub-giant (SGB) and 
red-giant (RGB) phases of these clusters.
They also detected the elusive, but predicted, bump along the RGB sequence.
Kerber \& Santiago, in these proceedings, have also tested different
stellar evolution models, with and without overshooting, using a sample
of 15 intermediate age clusters in the LMC.

Marigo \& Girardi (2007) have developed synthetic models to describe the
late evolutionary stage called thermally pulsating asymptotic giant branch 
(TP-AGB), where significant variability, mass loss and dredge-up 
events take place.
Their models were calibrated using the observed distribution of
both luminosities and lifetimes of carbon rich and oxygen rich 
(C and M, respectively) AGB stars in the two Magellanic Clouds.

Hennekemper et al (2008), again using HST/ACS, found a large number 
of pre-main sequence stars (PMS) in the young massive star forming 
region N66/NGC 346. Comparison of the observed PMS distribution in their
CMD with model predictions by Siess et al (2000) indicate the existence of
two recent episodes of star formation.
PMS stars, with $\tau \simeq 4$ Myrs, have also been recently detected 
by Carlson et al (2007) 
in the star forming region NGC 602, located in the SMC wing.

\section{Spectroscopy of MC clusters: a bit of kinematics and abundances}

Radial velocities and detailed abundance analysis of MC clusters require
use of spectroscopy. Recent detailed abundance studies of MC
clusters can be found in Trundle et al (2007), Hunter et al (2007)
and Mucciarelli et al (2008). The first two analyzed over 100 B stars 
in several star clusters in the Galaxy and in the MCs, and with
a large age span. They derived
atmospheric parameters and photospheric abundances of C, N, O, Mg, Si and Fe 
and investigated the effect of evolutionary processes such as rotation,
mass loss and binarism on the abundance of nitrogen.
 
Mucciarelli et al (2008) studied spectra of 27 red giant stars 
located in 4 rich and intermediate-age LMC clusters. Their analysis
yielded abundance ratios for about 20 atomic species, including $\alpha$, 
iron group and neutron capture elements. 

Kinematic studies of the MC cluster systems date from the 1980s. Freeman
et al (1983) analyzed a sample of 59 LMC clusters with individual radial
velocities accurate to 10-20 km/s. They concluded that the LMC cluster system
is consistent with a flatenned disk rotating at $\simeq 40$ km/s, although the 
disk geometry and systemic velocity was found to be different for young
and old clusters. Schommer et al (1992) used a larger cluster sample with 
more accurate velocities and concluded that all LMC clusters in their sample
have a single disk kinematics.
This result has been recently confirmed at $\simeq 2$ km/s precision 
by Grocholski et al (2006), who obtained 
metallicities and velocities from Ca II triplet lines for over 
200 stars in 28 populous LMC clusters observed with the Very Large Telescope 
(VLT). The authors also conclude that the LMC has no metallicity gradient.

For more on spectroscopic studies of star clusters in both Clouds, 
we refer to the contributions by
A. Ahumada, G. Bosch, A. Grocholski and A. Mucciarelli to this symposium.

\section{Comparison with other star cluster systems}

\begin{figure}[b]
\begin{center}
 \includegraphics[width=3.5in]{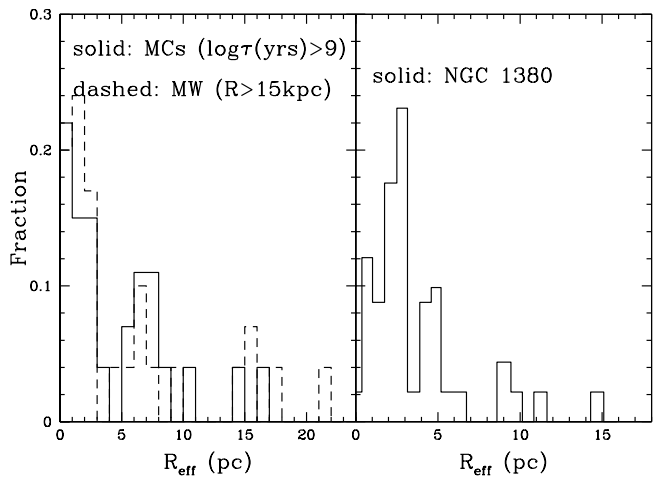} 
 \caption{Left panel: Distribution of half-light radii of clusters
in the Magellanic Clouds (solid) and in the Galaxy (dashed). Only clusters
with $log \tau(yrs) > 9$ in the Mackey et al (2003a,b) samples are 
included. The Galactic sample comes from Harris (1996) (and its 2003 
web update) and includes
only clusters farther than 15~kpc from the Galactic centre, in
order to avoid clusters more strongly affected by tidal effects.
Right panel: distribution of $R_{eff}$ for clusters in NGC 1380, which is
a luminous S0 galaxy in the Fornax cluster. The data are from Chies-Santos
et al (2007).}
 \label{mcrcdist}
\end{center}
\end{figure}

Knowledge on extragalactic clusters has increased immensely in the
last two decades. Among the most important breakthroughs are the 
universal (or nearly so) GCLF, the existence of bimodality 
in the distribution of
globular cluster colours and, more recently, the ability to measure 
cluster sizes. Correlations involving cluster luminosities, colours, 
sizes, location and host properties provide insight into the process 
of galaxy and cluster formation and evolution.

An interesting and relatively unexplored piece of information is the
size distribution of star clusters. The distributions of half-light radii
($R_{eff}$) of massive clusters in the Magellanic Clouds and in the Galaxy 
are shown in the left panel of Figure \ref{mcrcdist}. 
The figure reproduces Figure 2 from Mackey et al (2008). $R_c$ values
were transformed into $R_{eff}$ using the relation quoted by
Larsen (2001). For the Clouds, only clusters
older than $\tau~>~7$ Gyrs are shown, in order to make them more
comparable to their Galactic counterparts, which are all globular clusters. 
For the Galaxy, clusters
located closer than $R_{g} = 15$ pc to the centre were eliminated from the 
figure, as these suffer stronger
tidal effects, not occuring in the Clouds, and thus tend to be smaller
($R_{eff} \simeq 3$ pc). The two distributions are
very similar. The peak typical of Galactic globular clusters in the
inner halo is still the dominant one. Two other peaks, 
at $R_{eff} \simeq 7$ pc and at $R_{eff} \simeq 15$ pc are also seen.

The right panel shows the distribution of star clusters with measured
sizes in NGC 1380, which is a lenticular galaxy in the Fornax galaxy 
cluster. The data are from Chies-Santos et al (2007). There is no cut
in distance from the host centre, which likely explains the higher
fraction of clusters at $R_{eff} \simeq 3$ pc. Interestingly,
secondary peaks, similar to but at smaller radii than those in the Galaxy 
and in the Clouds, are also present.

\begin{figure}[b]
\begin{center}
 \includegraphics[width=4.0in]{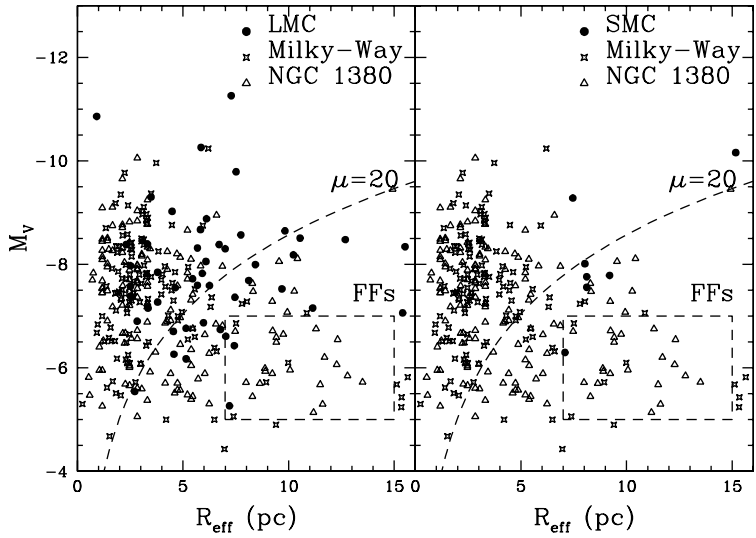} 
 \caption{Left panel: $M_V~vs.~R_{eff}$ relation for star clusters in
the LMC (solid circles), the Galaxy (stars) and in the luminous S0 in
Fornax galaxy clusters, NGC 1380 (triangles). Right panel: the same
as in the previous panel, but now the SMC clusters are compared to those
in the Galaxy and in NGC 1380. In both panels, the dashed box in the 
lower right is the
region occupied by faint fuzzies. The dashed line corresponds to a
surface brightness of $\mu_V$ = 20 mag arcsec$^{-2}$, where DSCs
preferentially lie.}
   \label{mcmvrc}
\end{center}
\end{figure}

The existence of clusters more extended than typical globulars
may be evidence of distinct populations of star clusters, possibly
with different formation and evolution histories. In well 
studied luminous 
early-type galaxies, which are the ones that harbour the richest 
cluster systems, extended clusters may also be distinguished
from the more common globulars in terms of other properties. 
Larsen \& Brodie (2000) and Brodie \&
Larsen (2002), for instance, found extended clusters with $7 \leq
R_{eff} \leq 15$ pc in two nearby lenticular galaxies. These were
named Faint Fuzzies, as they also tend to be underluminous compared
to the dominant cluster population. Their metallicities were estimated
espectroscopically, yielding $[Fe/H] \sim -0.6$. 
Finally, Burkert et al (2005) found evidence
that Faint Fuzzies in NGC 1023 are kinematically connected in a
ring-like structure.

Difuse star clusters (DSCs) may constitute another recently found
class. These were found by Peng et al (2006) in the ACS Virgo Cluster
survey (C\^ot\'e et al 2004). Several of luminous early-type galaxies, 
most of them S0s, 
have a large population of extended clusters of much lower surface brightness 
($\mu_g > 20$ mag~arcsec$^{-2}$) than typical globulars. The DSCs are also
redder than typical globular clusters, often having the same colour
as the difuse light from the host.

A useful tool used by Peng et al (2006) to separate different
types of clusters is the luminosity-size diagram. The $M_V - R_{eff}$
plane is shown in Figure \ref{mcmvrc}. Massive clusters in 3 markedly
different environments for cluster formation and evolution, the
Galaxy, the Magellanic Clouds and NGC 1380, are shown in the Figure. 
The data on $M_V$ are from the same references as the sizes. The dominant
population in the Galaxy and in NGC 1380 is made up of clusters
with $R_{eff} = 1-4$ pc and $-9 \leq M_V \leq -6$. Only a handful
of such globular-like clusters is found in the Clouds. The Clouds
also contribute little to the FFs box, shown in the lower right of
each panel. In fact, most of the clusters in the LMC and SMC tend to
follow the low-surface brightness line used by Peng et al (2006)
to characterize the DSCs. This means that the LMC and SMC have 
a large number of massive clusters that can be considered as
{\it structural counterparts} to DSCs, although they do not necessarily share
the other properties associated to these objects.

{\bf Acknowledgements}. I thank the SOC for the invitation to give
this review talk. C. Bonatto, L. Kerber and D. Mackey kindly provided
figures and data included in this review. Useful discussions preceding 
my conference presentation were carried out with colleagues in my home
institution, especially O. Kepler, M. Pastoriza, E. Bica and C. Bonatto.

\end{document}